\RequirePackage{fix-cm}
\documentclass[smallextended]{svjour3}       
\smartqed  
\usepackage{graphicx}
\def\gtap{\ \raise.3ex\hbox{$>$\kern-.75em\lower1ex\hbox{$\sim$}}\ }
\def\ltap{\ \raise.3ex\hbox{$<$\kern-.75em\lower1ex\hbox{$\sim$}}\ }
\begin{document}

\title{
Photo- and Electro-excitation of Bound Neutrons and Protons
}


\author{Satoshi X. Nakamura
}


\institute{
S.X. Nakamura \at
Laborat\'orio de F\'isica Te\'orica e Computacional-LFTC, 
Universidade Cruzeiro do Sul\\
 S\~ao Paulo, SP, 01506-000, Brazil\\
              \email{sxnakamura@gmail.com}           
}

\date{Received: date / Accepted: date}


\maketitle

\begin{abstract}
Data for pion photo-productions off the neutron ($\gamma n\to \pi N$)
have been primarily
extracted from deuteron-target data ($\gamma d\to \pi NN$)
by applying kinematical cuts
thereby isolating the quasi-free samples.  
We critically examine if the neutron-target data 
obtained through
this conventional procedure can be contaminated by final state
interactions (FSI) and/or the kinematical cuts. 
The analysis is conducted with a theoretical model for $\gamma d\to \pi NN$ 
that takes account of the impulse mechanisms supplemented by the $NN$
 and $\pi N$ rescattering mechanisms.
We show that the FSI effects still visibly remain in 
the extracted $\gamma$`$n$'$\to\pi N$
unpolarized cross sections and polarization asymmetries $E$
even after the kinematical cuts are applied.
We also find the FSI effects on $\gamma$`$n$'$\to\pi^0n$
can be somewhat different from those on $\gamma$`$p$'$\to\pi^0p$.
\end{abstract}

\section{Introduction}
\label{intro}

Meson production experiments with photon and electron beams have been very
active at facilities worldwide. 
A main interest is to gain information on the properties of nucleon
resonances ($N^*$) such as their pole positions 
and electromagnetic transition form factors,
using data complementary to those from the $\pi N$ scattering. 
Even the so-called over-complete measurement is planned to accurately
determine the amplitudes and thus $N^*$ properties with
significantly reduced model-dependence.
Lots of achievements towards this direction have been made, 
particularly with the free-proton target experiments,
and are summarized in this workshop~\cite{volker}. 

To gain a complete picture of the $N^*$ properties, we need not
only proton-target data but also neutron-target data. 
Measurements of the neutron-target data, including polarization
observables, have been also active recently. 
For example,
unpolarized differential cross sections ($d\sigma/d\Omega_\pi$)
and the polarization observable $E$ for $\gamma n\to\pi^0n$ 
have been measured at the MAMI~\cite{mami1,mami2},  
those for $\gamma n\to\pi^-p$ at the JLab~\cite{clas1,clas2},
and those for $\gamma n\to\eta n$ at the MAMI~\cite{mami3,mami4,mami5}.
More new (preliminary) results have been presented in this workshop~\cite{nstar}.

The primary interest in the neutron-target data is 
the electromagnetic neutron-to-$N^*$ transition form factors.
These $\gamma^{(*)}n\to N^*$ form factors are combined with
the $\gamma^{(*)}p\to N^*$ form factors to give the isospin structure of
the $\gamma^{(*)}N\to N^*$ form factors that are interesting quantities
for understanding the hadron structures.
The isospin decomposition is also necessary when we apply the form
factors to calculations of neutrino-induced meson productions~\cite{nuDCC}.
In addition to these primary interests,
the actual data have also brought unexpected surprises;
$\gamma n\to\eta n$ cross section data~\cite{mami3,graal,cbelsa}
revealed a narrow peak at $W\sim 1.68$~GeV 
($W$: the meson-baryon invariant mass)
which had not been found in the $\pi N$ and $\gamma p$ reaction data.

The deuteron has been primarily used in the experiments to extract the
neutron-target data.
Conventionally, a certain set of kinematical cuts is applied to the
deuteron data to supposedly isolate the quasi-free samples. 
However, this procedure has been always with a concern that
the kinematical cuts and/or nuclear effects such as final state
interactions (FSI) could distort the extracted neutron data from the
true free neutron-target observables.
The purpose of this work is to critically examine 
within a theoretical model 
the effects of the FSI and 
kinematical cuts on (un)polarized $\gamma n\to\pi N$ 
cross sections
extracted from $\gamma d\to\pi NN$ cross sections.

\section{Model}
\label{sec:model}

\begin{figure}
  \includegraphics[width=1\textwidth]{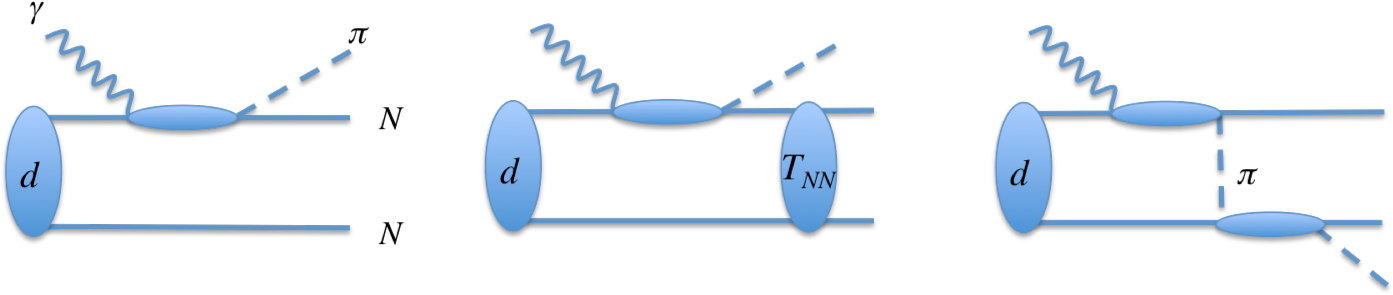}
\caption{
Diagrammatic representation of reaction mechanisms considered in this work for $\gamma d\to\pi N N$:
(left) impulse, (center) $NN$ rescattering, 
(right) $\pi N$ rescattering mechanisms.
}
\label{fig:diag}
\end{figure}

This study will be based on a $\gamma d\to\pi NN$ reaction model that
consists of the impulse [Fig.~\ref{fig:diag}(left)],
$NN$ rescattering [Fig.~\ref{fig:diag}(center)], and
$\pi N$ rescattering [Fig.~\ref{fig:diag}(right)] mechanisms.
The model needs to be incorporated with realistic elementary amplitudes for
the subprocesses involved.
Regarding $\gamma N \to \pi N$ and $\pi N \to \pi N$ amplitudes,
we employ those generated with a dynamical coupled-channels (DCC)
model~\cite{knls13,knls16}.
The DCC model takes account of coupled-channels relevant to the
nucleon resonance region, and 
has been developed through analyzing 
$\sim 27,000$ data points of $\pi N, \gamma N \to \pi N, \eta N, K\Lambda, K\Sigma$ 
from the thresholds up to $W\ltap 2.1$~GeV.
As for the deuteron wave function and the $NN\to NN$ amplitudes, 
we employ those generated with the CD-Bonn potential~\cite{cdbonn}.

\section{Results 
\small{(all numerical results are preliminary)}}
\label{sec:result}

\begin{figure}
  \includegraphics[width=0.52\textwidth]{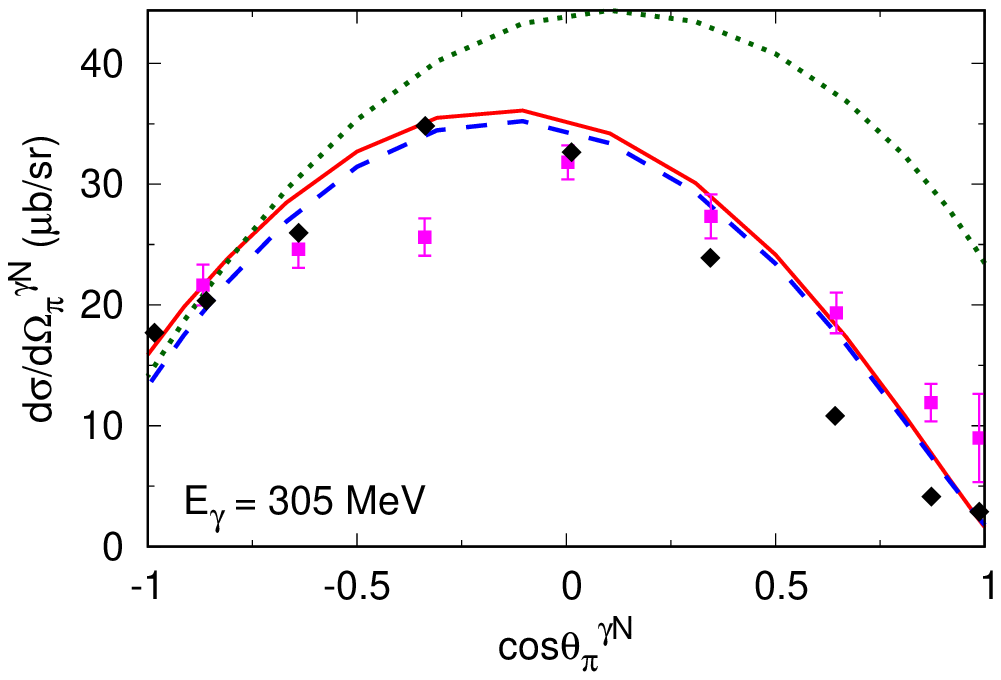}
\hspace{-3mm}
  \includegraphics[width=0.52\textwidth]{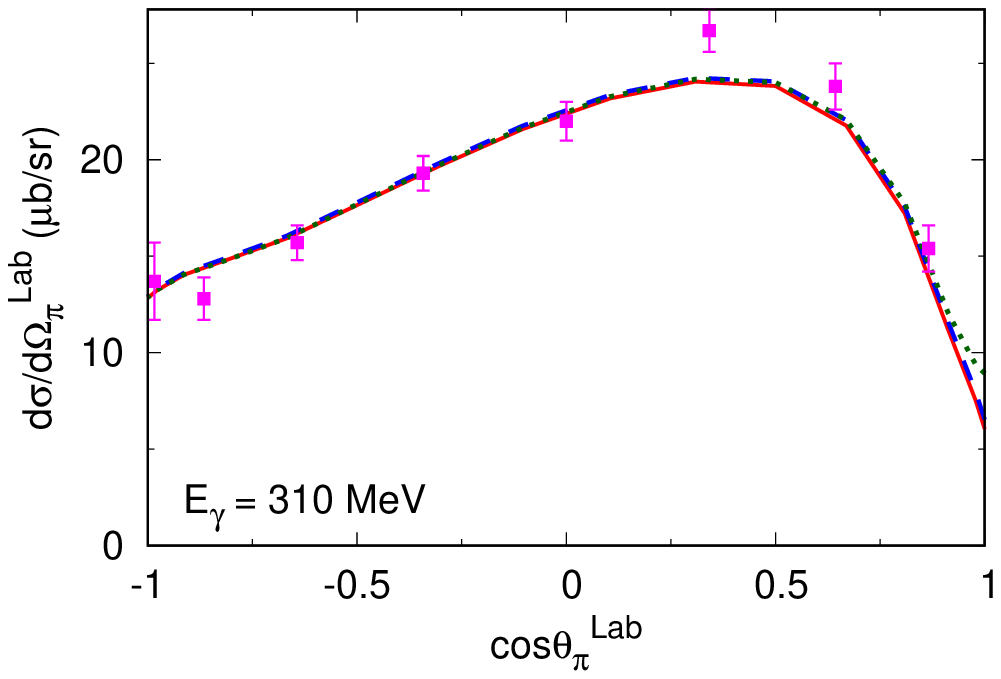}
\caption{
The pion angular distribution for 
$\gamma d\to \pi^0 pn$ (left)
and $\gamma d\to \pi^- pp$ (right).
The photon energy in the laboratory frame is indicated in the figure.
The curves are obtained with the impulse approximation (green dotted),
the impulse and $NN$ rescattering mechanisms (blue dashed), and the
full model (red solid) that also includes the $\pi N$ rescattering mechanism.
The data are from 
Ref.~\cite{gd-pi0pn_data} (magenta squares)
and Ref.~\cite{gd-pi0pn_data2} (black diamonds)
for $\gamma d\to \pi^0 pn$, 
and from Ref.~\cite{gd-pimpp_data} for 
$\gamma d\to \pi^- pp$.
}
\label{fig:1}
\end{figure}

First we confront parameter-free
 model predictions for $\gamma d\to \pi NN$ cross
sections with data
to examine the soundness of the model and FSI effects. 
Here, 'parameter-free' means that we did not adjust any model parameters
using $\gamma d\to \pi NN$ cross section data.
The comparisons are presented in Fig.~\ref{fig:1}.
The agreement with data is reasonably good overall.
For $\gamma d\to \pi^0 pn$,
a large reduction of the cross sections
due to the $NN$ rescattering is essential for the good agreement with
the data.
This reduction is mainly caused by the orthogonality between the deuteron wave
function and the $^3S_1$ partial wave in the final $NN$ state. 
The $\pi N$ rescattering effect is rather moderate.
Meanwhile, for $\gamma d\to \pi^- pp$ where the orthogonality mentioned
above does not come into play, the FSI effects are rather small. 
We mention that the DCC-based deuteron reaction model predicts
$\gamma d\to \eta pn$ cross sections that are in excellent agreement
with data~\cite{etaN}.

\begin{figure}[b]
  \includegraphics[width=0.52\textwidth]{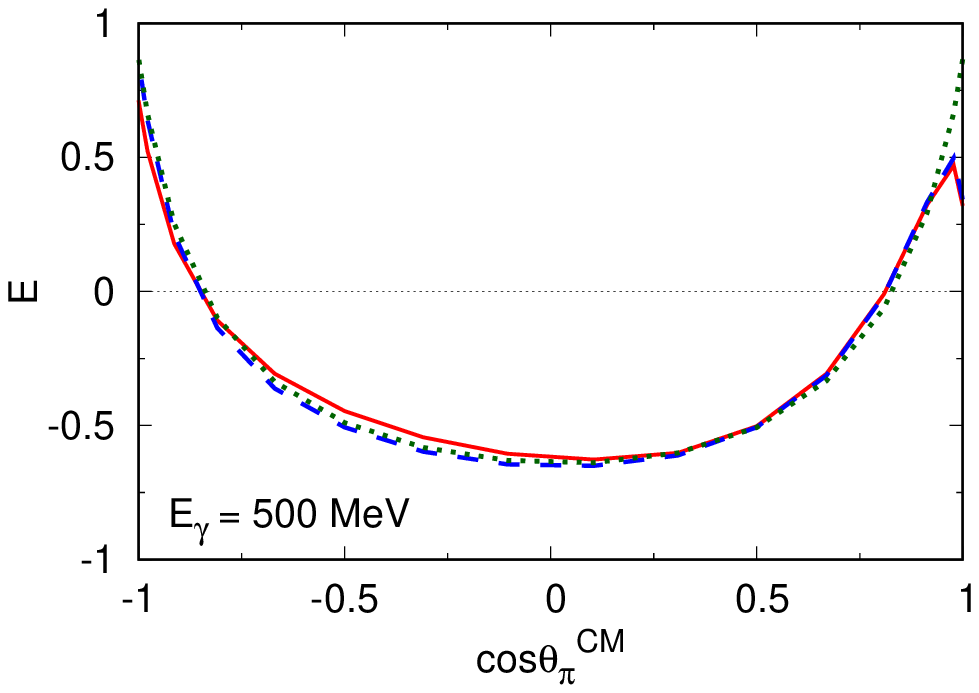}
\hspace{-3mm}
  \includegraphics[width=0.52\textwidth]{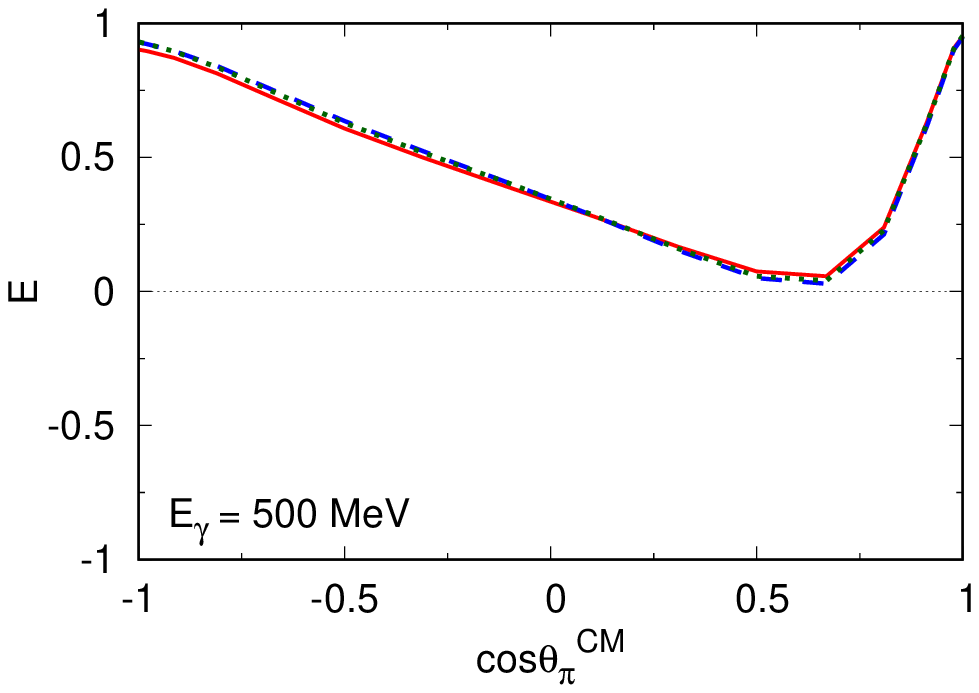}
\caption{
The polarization asymmetry $E$ for
$\gamma d\to \pi^0 pn$ (left)
and $\gamma d\to \pi^- pp$ (right).
The other features are the same as those in Fig.~\ref{fig:1}.
}
\label{fig:2}
\end{figure}

We also present in Fig.~\ref{fig:2} the polarization asymmetry $E$
defined by 
$E = (\sigma_{+-}-\sigma_{++})/(\sigma_{+-}+\sigma_{++})$,
where $\sigma_{++}$ ($\sigma_{+-}$) is the $\gamma d\to\pi NN$
cross section for which
the photon circular polarization and the deuteron spin orientation are
parallel (antiparallel).
Comparing Figs.~\ref{fig:1} and \ref{fig:2},
we find that the FSI effects on the $\gamma d\to\pi^0pn$
cross sections are significantly canceled
in the ratio $E$.

Now we extract cross sections (including polarization observables) for
quasi-free $\gamma n\to\pi N$ 
from those for $\gamma d\to\pi NN$ in a conventional manner.
A set of kinematical cuts is
applied to the $\gamma d\to\pi NN$
cross sections generated with
the DCC-based model, and
then further correction is
made for the Fermi motion
(Fermi-unsmearing).
By comparing the extracted quasi-free cross sections
with the corresponding free ones, which are calculated with
the same elementary amplitudes used in the $\gamma d\to\pi NN$ model,
we address the questions on 
how the extracted cross sections could be distorted by
the FSI and kinematical cuts.
We employ the kinematical cuts that have been used in recent
experimental analyses
to extract $\gamma n\to\pi^-p$ from $\gamma d\to\pi^-pp$,
as summarized in Table~\ref{tab:1}.
\begin{table}[t]
\begin{center}
\begin{tabular}[t]{c|cc}\hline
& $d\sigma/d\Omega_\pi$~\cite{clas2} & $E$~\cite{clas1}  \\\hline
$\pi^-$ momentum (GeV)& $>0.1$ & $>0.4$  \\
Faster proton momentum (GeV)& $>0.36$ & $>0.4$ \\
Slower proton momentum (GeV)& $<0.2$ & $<0.1$  \\
$\Delta\phi=|\phi_{\pi^-} - \phi_{\rm faster\ proton}|$
& - & $160^\circ < \Delta\phi < 200^\circ$  \\\hline
\end{tabular}
\end{center} 
\caption{The kinematical cuts used in the experimental
analyses~\cite{clas1,clas2} for extracting unpolarized cross sections
($d\sigma/d\Omega_\pi$) and the polarization asymmetry $E$ for 
$\gamma n\to \pi^-p$ 
from those for  $\gamma d\to \pi^-pp$.
The azimuthal angle difference between $\pi^-$ and the faster proton
is denoted by $\Delta\phi$.
}
\label{tab:1}
\end{table}
We also use the same cuts to extract $\gamma n\to\pi^0 n$ from 
$\gamma d\to\pi^0pn$.
For the Fermi-unsmearing, we follow the procedure described in 
Appendix~B of Ref.~\cite{tarasov2016}.

We show in Fig.~\ref{fig:3} the quasi-free $\gamma n\to \pi^0 n$ cross sections
extracted from theoretical $\gamma d\to\pi^0pn$ cross section,
using the kinematical cuts of Table~\ref{tab:1} and the Fermi unsmearing. 
The green triangles (blue circles) [red squares] in the figure are extracted
from the $\gamma d\to \pi^0 pn$ cross sections
calculated with the impulse approximation
(the impulse and $NN$ rescattering mechanisms)
[the full model].
\begin{figure}
  \includegraphics[width=0.52\textwidth]{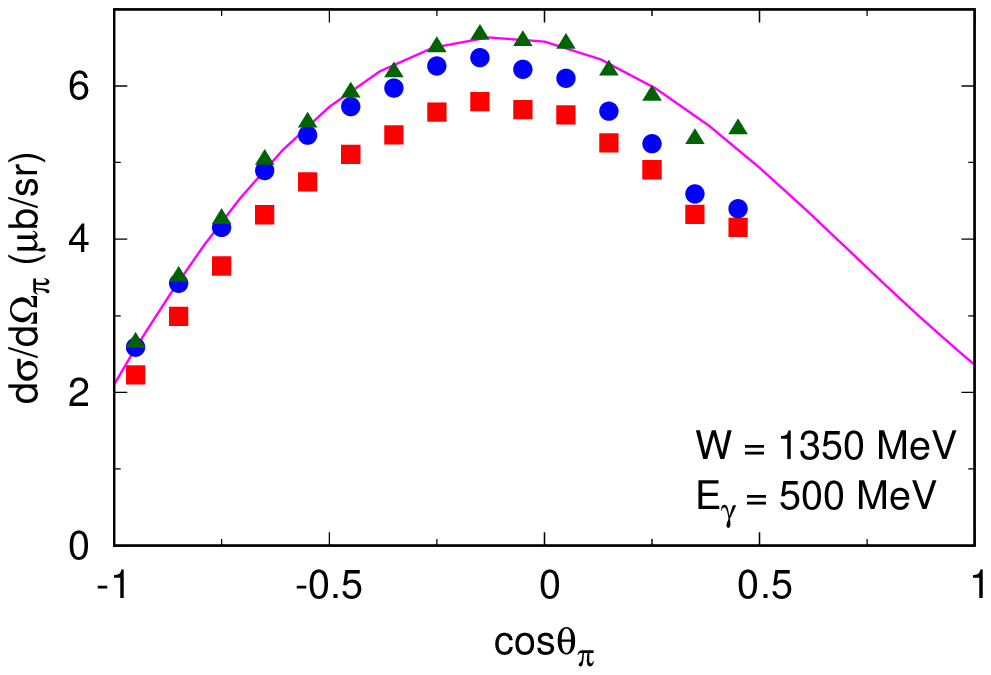}
\hspace{-3mm}
  \includegraphics[width=0.52\textwidth]{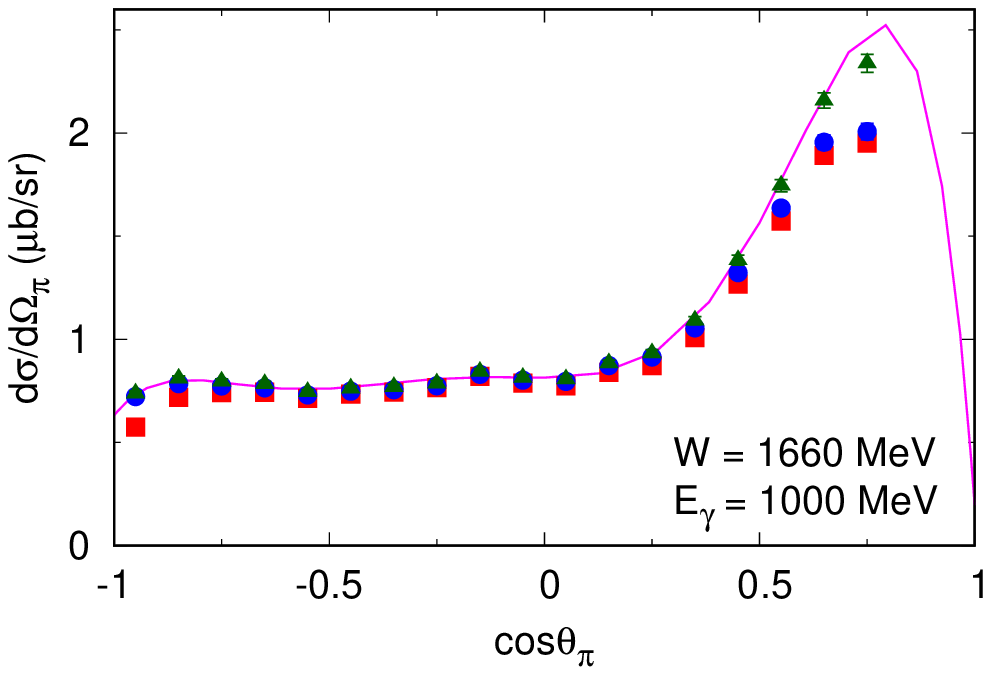}
\caption{
The pion angular distribution for $\gamma n\to \pi^0n$ 
extracted from $\gamma d\to \pi^0 pn$.
The photon energy ($E_\gamma$)
in the laboratory frame and the final $\pi^0n$ invariant mass ($W$)
for $\gamma d\to \pi^0 pn$ are indicated in the figure.
The points with error bars are extracted 
from theoretical $\gamma d\to \pi^0 pn$ differential cross sections,
using the kinematical cuts of Table~\ref{tab:1} and the Fermi
unsmearing. 
When the theoretical $\gamma d\to \pi^0 pn$ cross sections
are calculated with the impulse approximation
(the impulse and $NN$ rescattering mechanisms)
[the full model],
the green triangles (blue circles) [red squares] are obtained.
%
%
The errors are only statistical from the Monte-Carlo integral, and are
 not shown when smaller than the point size.
The solid curve is the free $\gamma n\to \pi^0n$ cross sections at $W$
from the DCC model.
}
\label{fig:3}
\end{figure}
The phase-space integral has been done with the Monte-Carlo method, and
thus the numerical results are given by the points with error bars; the
errors include only statistical ones associated with the Monte-Carlo method. 
The kinematical cuts remove the forward $\pi$ kinematical regions.
We can see that a significant reduction due to the FSI remains even
after the kinematical cuts have been applied. 
The $NN$ and $\pi N$ FSI contributions are comparably visible.
Therefore, only the quasi-free cross sections 
extracted from $\gamma d\to \pi^0 pn$ 
calculated with the impulse approximation reproduce well the free cross
sections.

\begin{figure}
  \includegraphics[width=0.52\textwidth]{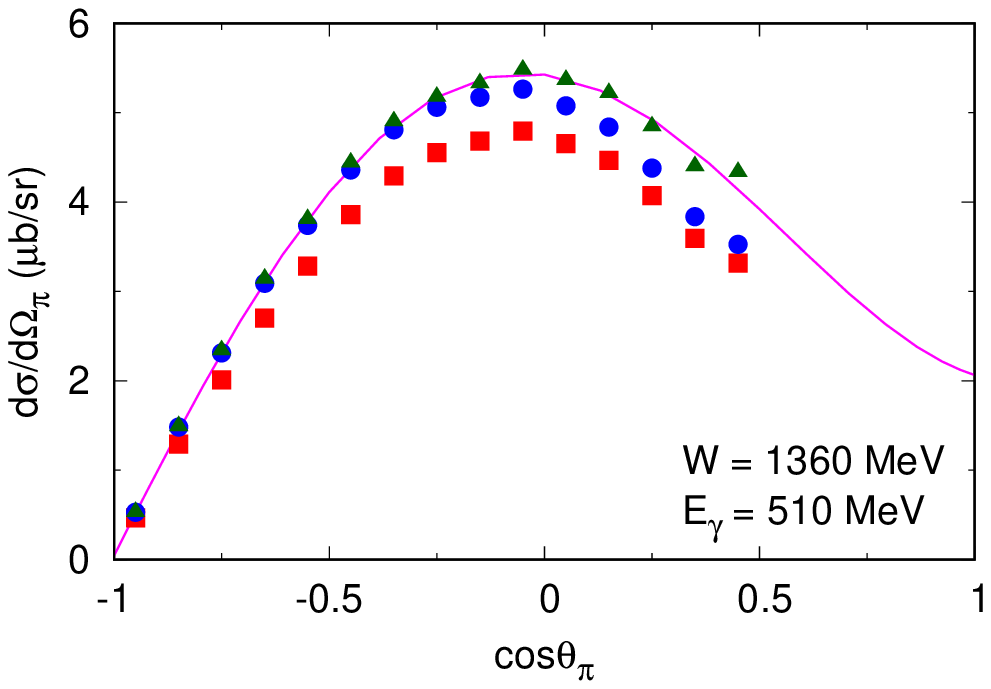}
\hspace{-3mm}
  \includegraphics[width=0.52\textwidth]{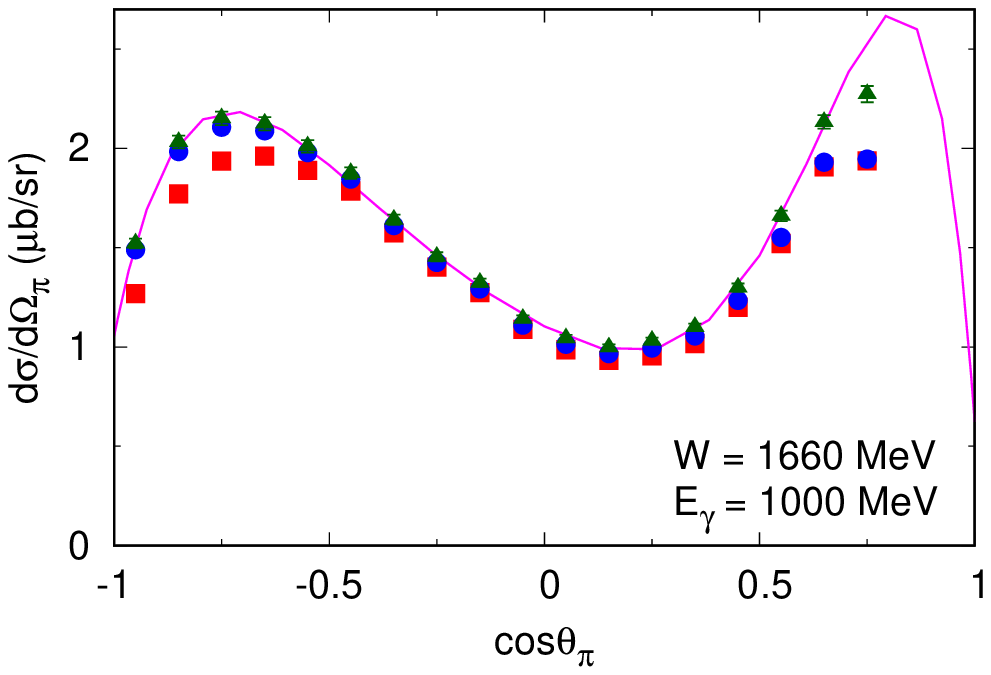}
\caption{
The pion angular distribution for $\gamma p\to \pi^0p$ 
extracted from $\gamma d\to \pi^0 pn$.
The other features are the same as those in Fig.~\ref{fig:3}.
}
\label{fig:4}
\end{figure}

It would be interesting to compare the FSI effects on the quasi-free
$\gamma n\to \pi^0 n$ and those on the $\gamma p\to \pi^0 p$ cross sections.
This is particularly important because an experimental analysis sometimes assumes
that the FSI effects are the same for both, as in Ref.~\cite{mami1}.
Therefore, we show in Fig.~\ref{fig:4} the quasi-free $\gamma p\to \pi^0 p$ 
cross sections.
By comparing Figs.~\ref{fig:3} and \ref{fig:4}, the FSI effects
are somewhat different between $\gamma n\to\pi^0 n$
and $\gamma p\to\pi^0 p$ cross sections; generally a few percents difference in the
reduction rates.

\begin{figure}
  \includegraphics[width=0.52\textwidth]{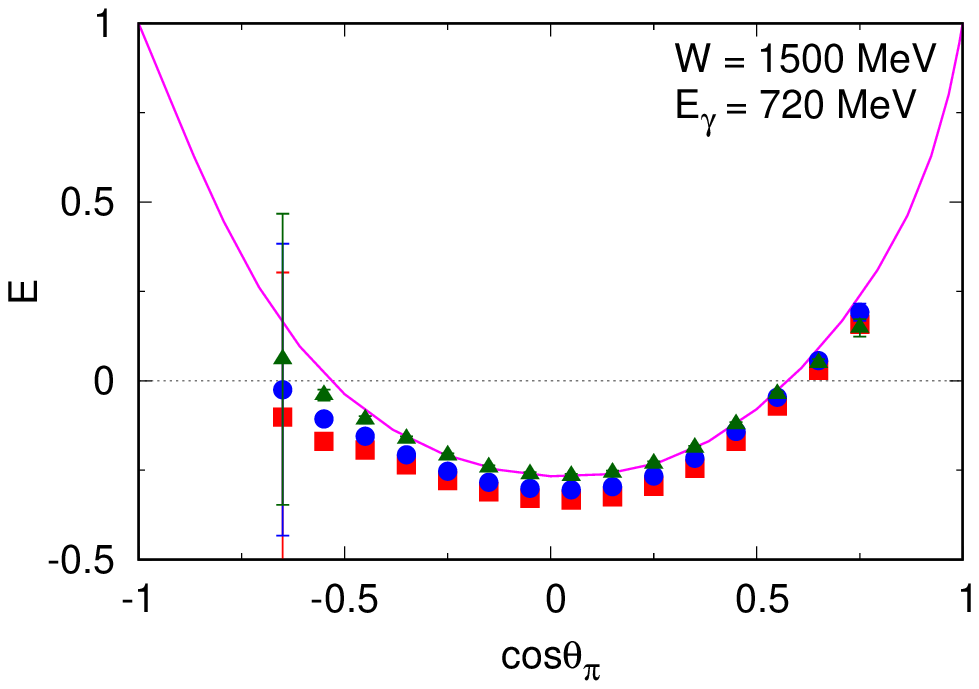}
\hspace{-3mm}
  \includegraphics[width=0.52\textwidth]{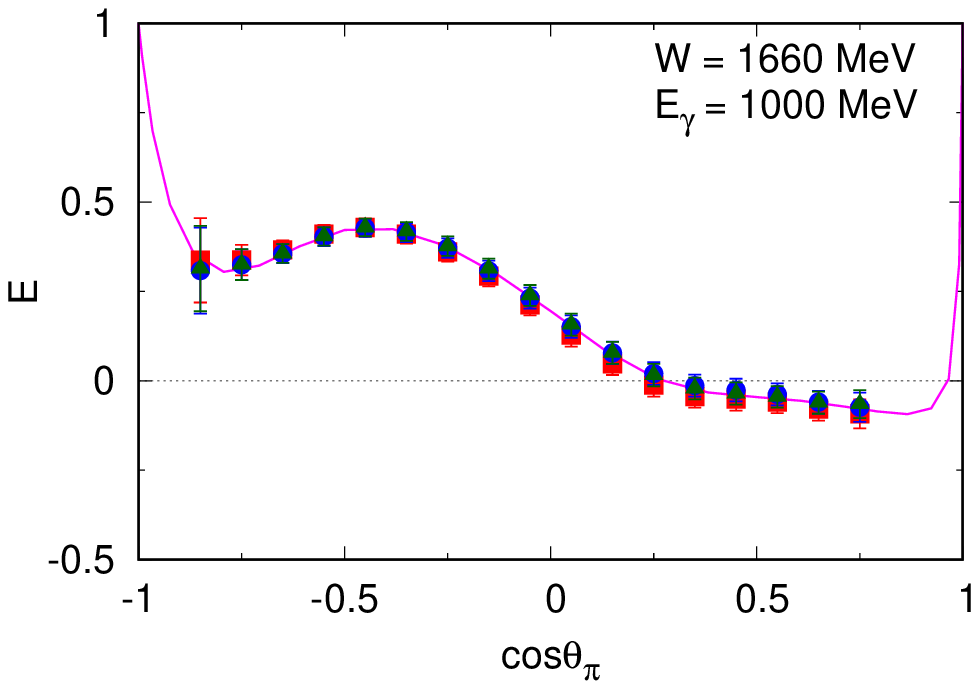}
\caption{
The polarization observable $E$ for $\gamma n\to \pi^0n$ 
extracted from $\gamma d\to \pi^0 pn$.
The other features are the same as those in Fig.~\ref{fig:3}.
}
\label{fig:5}
\end{figure}

Next we discuss the polarization observable $E$ for the 
quasi-free $\gamma n\to \pi^0 n$.
This polarization asymmetry is defined by 
$ E = (\sigma_{1/2}-\sigma_{3/2})/(\sigma_{1/2}+\sigma_{3/2})$,
where $\sigma_{3/2}$ ($\sigma_{1/2}$) is the $\gamma N$ cross section for which
the photon circular polarization and the nucleon spin orientation are
parallel (antiparallel).
We show in Fig.~\ref{fig:5} $E$ for the quasi-free 
$\gamma n\to \pi^0 n$.
The magnitude of the FSI effects on $E$ are smaller than those on
$d\sigma/d\Omega_\pi$ because of the partial cancellation in the ratio.
The previous experimental analysis~\cite{mami2} assumed no
FSI effects on $E$, invoking the cancellation.
However, we can still see nonnegligible FSI effects on $E$ for 
$\gamma n\to \pi^0 n$ at some $W$, and thus
depending on the precision of data, corrections from this effect would be
needed.

\begin{figure}
  \includegraphics[width=0.52\textwidth]{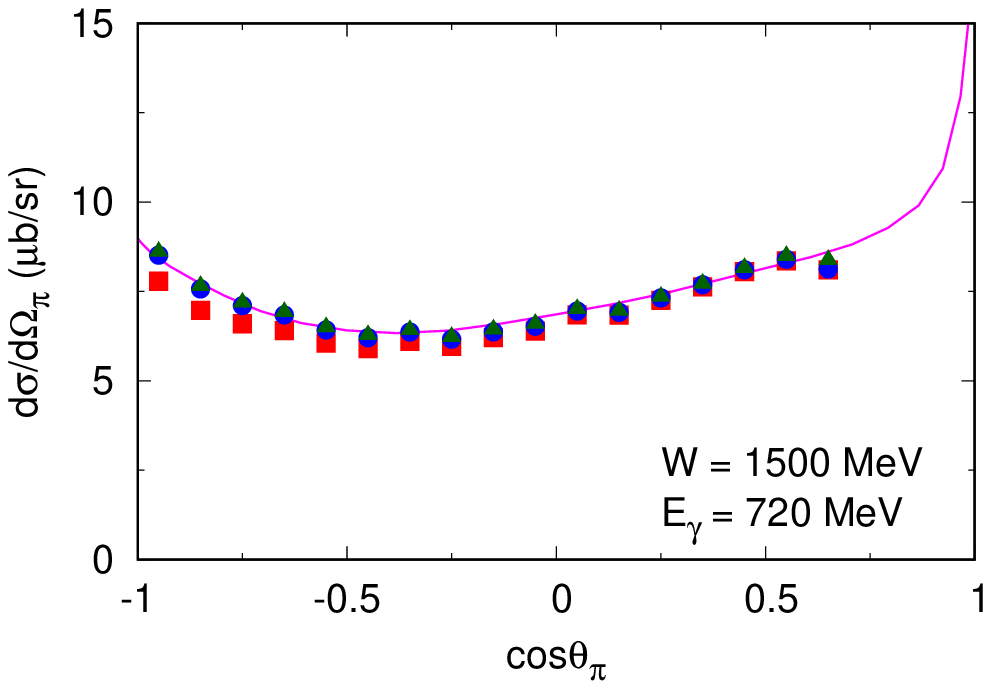}
\hspace{-3mm}
  \includegraphics[width=0.52\textwidth]{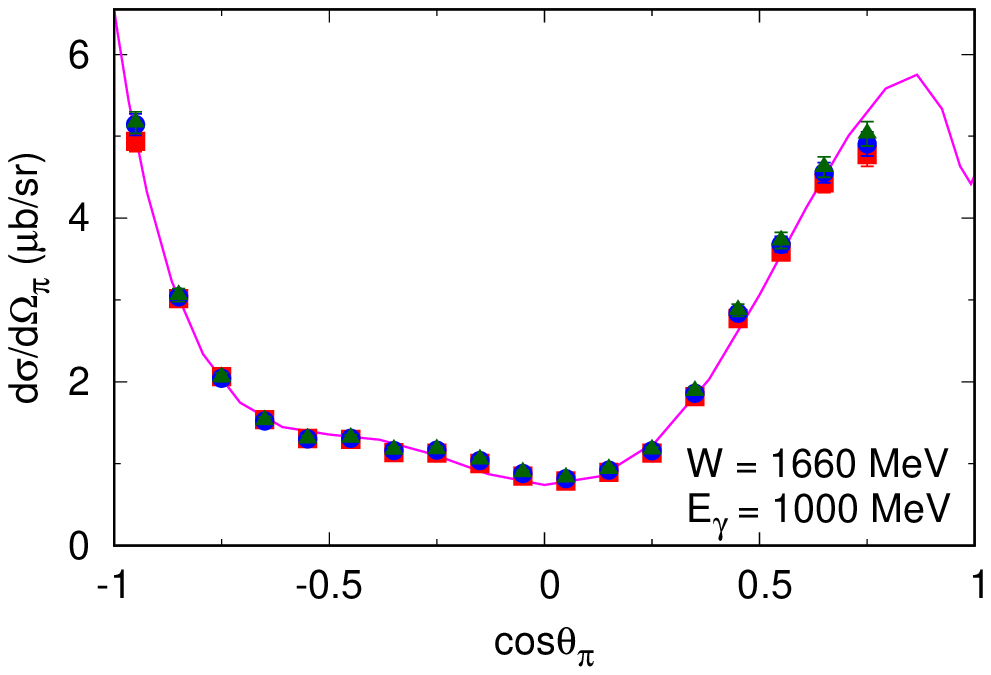}
  \includegraphics[width=0.52\textwidth]{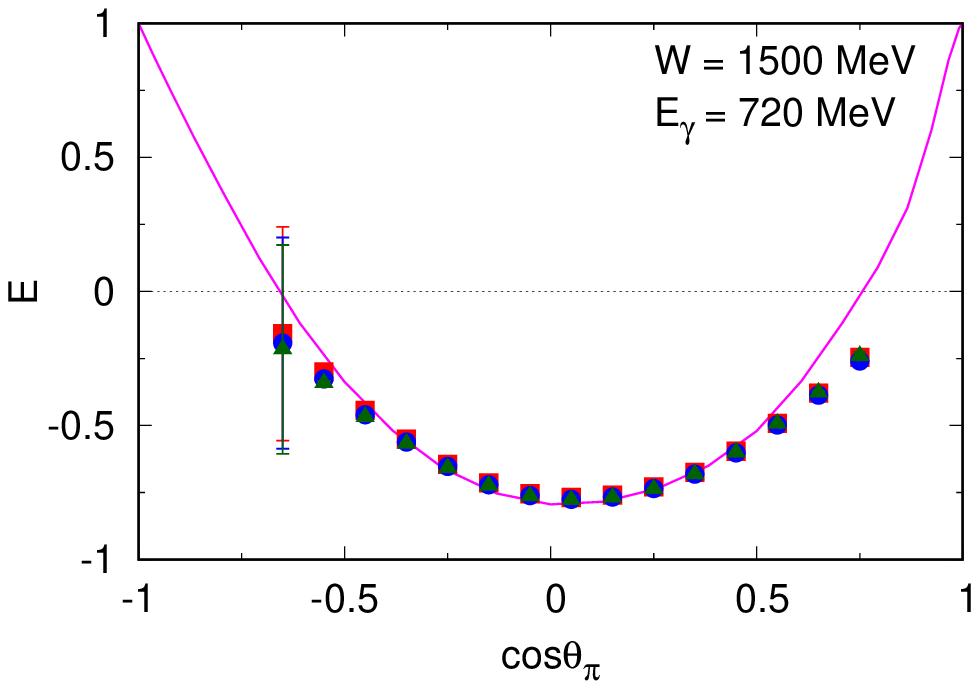}
\hspace{-3mm}
  \includegraphics[width=0.52\textwidth]{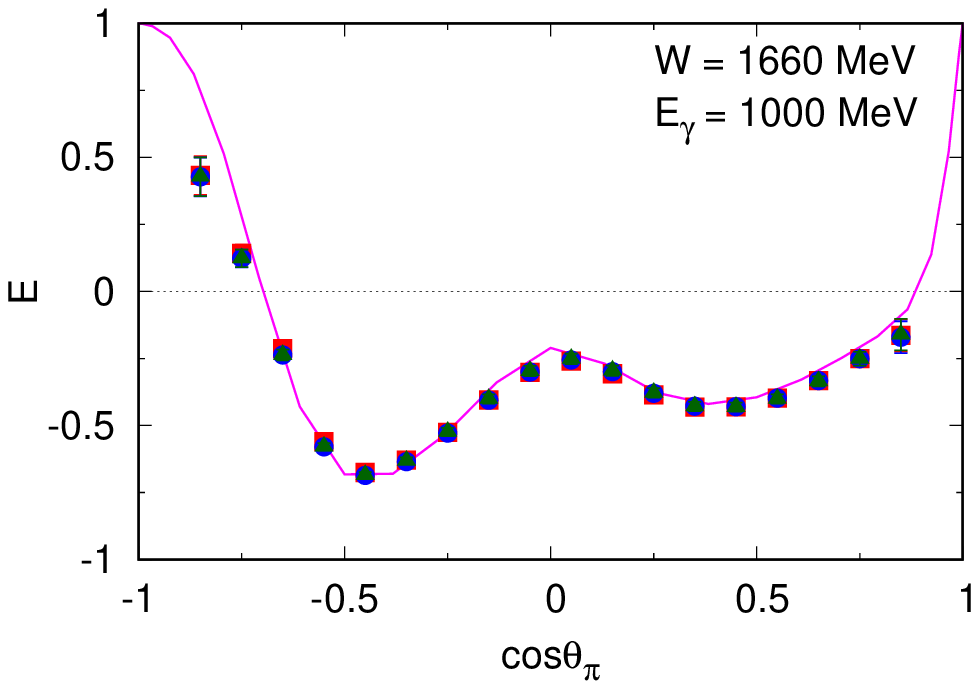}
\caption{
The $d\sigma/d\Omega_\pi$ and $E$ for $\gamma n\to \pi^-p$ 
extracted from $\gamma d\to \pi^- pp$.
The other features are the same as those in Fig.~\ref{fig:3}.
}
\label{fig:6}
\end{figure}

Finally, we study the FSI effects on 
$d\sigma/d\Omega_\pi$ and $E$ for the quasi-free $\gamma n\to\pi^-p$
extracted from $\gamma d\to\pi^-pp$.
Our numerical result is give in Fig.~\ref{fig:6}.
We again find that the quasi-free cross sections 
extracted from $\gamma d\to\pi^-pp$ in the impulse approximation
reproduce well the free cross sections. 
For the backward pion kinematics, we find $\sim 9$\% reduction of
$d\sigma/d\Omega_\pi$ at $W=1500$~MeV due to the $\pi N$ rescattering. 
Tarasov et al. also reported a similar finding~\cite{tarasov2011}.
Regarding the polarization asymmetry $E$, we do not find a visible FSI
effect. 
However, we find 
that the extracted quasi-free $E$ at the kinematical ends
are noticeably different from the free $E$.
This might be an artifact of the kinematical cuts, and a more elaborate
study is underway.

As seen in Figs.~\ref{fig:3}-\ref{fig:6}, 
the FSI effects often shift the extracted $\gamma n\to\pi N$
observables from the corresponding free ones. 
This may seem to indicate that 
we cannot correctly extract $\gamma n\to\pi N$
observables using the conventional method based on the kinematical cuts.
Meanwhile, it is not computationally practical to determine the 
$\gamma n\to\pi N$ amplitudes (and thus the observables) by directly fitting them to
the $\gamma d\to\pi NN$ data through
a reaction model such as the one used in this work.
A possible option would be to take an iterative procedure as follows:
\begin{enumerate}
\item 
Start with a certain set of parameters of a dynamical model that generates
$\gamma n\rightarrow \pi N$ amplitudes.
Implement the amplitudes into a dynamical 
$\gamma d\to\pi NN$ reaction model.
\item With the same set of kinematical cuts as used in an experiment, 
calculate $\gamma d\to\pi NN$ cross sections
with the dynamical model including the FSI,
and extract the $\gamma n\to\pi N$ cross sections 
using the conventional method.
By comparing the extracted $\gamma n\rightarrow \pi N$ cross sections
with the free ones from the $\gamma n\rightarrow \pi N$ amplitudes built
      in the $\gamma d$ model, the FSI correction factor on the extracted cross
      sections is obtained.
\item 
Extract $\gamma n\rightarrow \pi N$ data from 
 $\gamma d\to\pi NN$ experimental data using the conventional method.
The obtained quantity is further 
multiplied by the FSI
correction factor estimated in
the previous step,
then $\gamma n\rightarrow \pi N$ cross section data are extracted.
\item 
Calculate $\gamma n\rightarrow \pi N$ cross sections with the elementary
amplitudes that have been built in the $\gamma d$ model, and
compare them with the extracted data.
If they agree, the extraction of the $\gamma n\rightarrow \pi N$ data
as well as that of the corresponding amplitude are complete.
If not, fit the extracted data 
with the dynamical model for $\gamma n\rightarrow \pi N$
by adjusting their parameters, 
then return to the step 1.
\end{enumerate}

\begin{acknowledgements}
The author acknowledges H. Kamano, T.-S.H. Lee and T. Sato for their collaborations.
This work was supported in part by 
Funda\c{c}\~ao de Amparo \`a Pesquisa do Estado de S\~ao Paulo-FAPESP,
Process No.~2016/15618-8.
Numerical computations in this work were carried out
with SR16000 at YITP in Kyoto University,
the High Performance Computing system at RCNP in Osaka University,
the National Energy Research Scientific Computing Center, which is
supported by the Office of Science of the U.S. Department of Energy
under Contract No. DE-AC02-05CH11231, and resources provided on Blues
and/or Bebop, high-performance computing cluster operated by the
Laboratory Computing Resource Center at Argonne National Laboratory.

\end{acknowledgements}



\end{document}